# THREE-PORT IMPEDANCE MODEL AND VALIDATION OF VSCs FOR STABILITY ANALYSIS

*Haoxiang Zong[1*], Chen Zhang[1], Xu Cai[1], Marta Molinas[2]*

[1]*Wind Power Research centre, Shanghai Jiao Tong University, Shanghai, China*
[2]*Department of Engineering Cybernetics, Norwegian University of Science and Technology, Trondheim, Norway*
*\*haoxiangzong@sjtu.edu.cn*

## Abstract

Modern power system is undergoing a paradigm shift from the synchronous generators-based system to the power electronics converters-dominated system. With the high penetration of converters, serious stability problems are provoked, especially the wideband oscillations (e.g., sub-synchronous oscillations, harmonic oscillations, etc). Various studies have been conducted in this respect, while most of them separate the ac-side stability with the dc-side stability. However, for the stability analysis of the hybrid AC/DC grid, it is necessary to consider the converter's ac-side and dc-side, simultaneously. In this paper, the stability analysis of voltage source converters (VSCs) considering both ac and dc dynamics is carried out. At first, the three-port AC/DC admittance model of VSCs is established, and the corresponding measurement method from simulations is presented to validate its accuracy. Secondly, based on such three-port model, two stability analysis methods are presented: the one is based on the system open-loop model, where the stability can be judged via the Generalized Nyquist Criterion (GNC); the other one is based on the system closed-loop model, whose stability can be predicted through the pole-zero calculation. At last, a test AC/DC system is built in MATLAB/Simulink, by which the effectiveness of the three-port model-based stability analysis is validated.

## 1 Introduction

Nowadays, the power system is gradually transformed from pure AC or DC grid into the hybrid AC/DC grid [1], mainly formed by power electronics converters-dominated renewable energy sources. Unlike synchronous generators, the power electronic converters, e.g., voltage source converters (VSCs), exhibit unique characteristics of wideband dynamics and AC/DC couplings, imposing great challenges on the stable operation of the power system [2].

In this respect, the impedance-based stability tools [3] are favoured and numerous studies aimed at characterizing small-signal behaviours of VSCs have been carried out [4]-[6]. For the ac-side stability, the frequency couplings [3] inside the ac port are considered, from which various 2×2 impedance models emerged including *dq* models [4], complex vector models [5], etc. It has been proved that the accuracy of 2×2 models is enough for the converters' ac-side stability analysis under symmetrical conditions [5]. For the dc-side stability analysis, the one-dimensional impedance model [6] is adopted to formulate the dc network model, whose stability can be studied via impedance aggregation methods [7].

To further improve the stability analysis accuracy, the influence of ac-/dc- side dynamics on the opposite side is noticed by some scholars and an equivalent ac-/dc- side impedance model [8], [9] is proposed. For example, the Ref [8] incorporate the dc-link dynamics of the Type-III wind turbine system into its ac-side impedance model through an indicator function; the Ref [9] proposed a building block-based modelling method for the Type-IV wind turbine, based on which the ac-side dynamics can be incorporated into the dc-side impedance model. However, neither the ac-side equivalent model nor the dc-side equivalent model is convenient for the stability analysis of large-scale AC/DC interconnected system [10]. To address this issue, this paper presents a three-port AC/DC admittance model so as to consider converters' AC/DC dynamics, simultaneously. Furthermore, the corresponding measurement method from simulations is also provided. Based on such three-port admittance model, two stability analysis methods, i.e., open-loop method and closed-loop method, are discussed and applied in the converter AC/DC stability evaluation.

The rest of the paper is arranged as follows. Section 2 provides the VSC three-port AC/DC admittance model; Section 3 presents the frequency scanning method and validate the accuracy of the established model; Section 4 demonstrates the open-loop and closed-loop method in the AC/DC stability analysis; Section 5 concludes the paper.

## 2. Three-port AC/DC Admittance Model

In this section, the three-port admittance model of the VSC is built in a step-by-step manner. The topology of the VSC as well as its control scheme is presented in Fig. 1.

*1.1 Main circuit modelling*

As shown in Fig. 1, the *d-q* small-signal admittance model of the ac-side filter can be derived as:

$$\Delta \boldsymbol{u}_{\mathrm{c}dq} = -\boldsymbol{Z}_{dq}^{\mathrm{f}}(s)\Delta \boldsymbol{i}_{\mathrm{g}dq} + \Delta \boldsymbol{u}_{\mathrm{g}dq} \qquad (1)$$

where $\Delta \boldsymbol{u}_{cdq} = \begin{bmatrix} \Delta u_{cd} \\ \Delta u_{cq} \end{bmatrix}$, $\boldsymbol{Z}_{dq}^{f}(s) = \begin{bmatrix} R_f + sL_f & -\omega_1 L_f \\ \omega_1 L_f & R_f + sL_f \end{bmatrix}$, $\Delta \boldsymbol{i}_{gdq} = \begin{bmatrix} \Delta i_{gd} \\ \Delta i_{gq} \end{bmatrix}$ and $\Delta \boldsymbol{u}_{gdq} = \begin{bmatrix} \Delta u_{gd} \\ \Delta u_{gq} \end{bmatrix}$.

Since the VSC's ac-side is coupled with its dc-side via the power balance, the ac-side and dc-side dynamics can be represented in (2) and (3), respectively.

$$\Delta \boldsymbol{u}_{cdq} = \Delta \boldsymbol{m}_{dq}(s) V_{dc0} + \boldsymbol{m}_{dq0} \Delta u_{dc} \qquad (2)$$

$$\Delta i_{dc} = \frac{3}{2V_{dc0}} \left( \boldsymbol{U}_{cdq0}^{T} \Delta \boldsymbol{i}_{gdq} + \boldsymbol{I}_{gdq0}^{T} \Delta \boldsymbol{u}_{cdq} \right) - \frac{I_{dc0}}{V_{dc0}} \Delta u_{dc} \qquad (3)$$

where $\Delta \boldsymbol{m}_{dq}(s) = \begin{bmatrix} \Delta m_d \\ \Delta m_q \end{bmatrix}$, $\boldsymbol{m}_{dq0} = \begin{bmatrix} m_{d0} \\ m_{q0} \end{bmatrix}$; $\boldsymbol{U}_{cdq0}^{T} = \begin{bmatrix} U_{cd0} & U_{cq0} \end{bmatrix}$, $\boldsymbol{I}_{gdq0}^{T} = \begin{bmatrix} I_{gd0} & I_{gq0} \end{bmatrix}$

*1.2 Control part modelling*

*1.2.1 PLL*

The small-signal model of the phase locked loop (PLL) is like:

$$\Delta \theta_{pll} = \underbrace{\frac{H_{pll}(s)}{s + U_{gd0} \cdot H_{pll}(s)}}_{T_{pll}(s)} \Delta u_{gq} \qquad (4)$$

The relationship between the converter domain and the system domain is like:

$$\Delta \boldsymbol{i}_{gdq}^{c} = \Delta \boldsymbol{i}_{gdq} + \boldsymbol{I}_{gdq0}^{pll}(s) \Delta \boldsymbol{u}_{gdq}$$
$$\Delta \boldsymbol{u}_{gdq}^{c} = \boldsymbol{U}_{gdq0}^{pll}(s) \Delta \boldsymbol{u}_{gdq} \qquad (5)$$
$$\Delta \boldsymbol{m}_{dq}^{c} = \Delta \boldsymbol{m}_{dq} + \boldsymbol{M}_{dq0}^{pll}(s) \Delta \boldsymbol{u}_{gdq}$$

where $\Delta \boldsymbol{i}_{gdq}^{c}$, $\Delta \boldsymbol{u}_{gdq}^{c}$ and $\Delta \boldsymbol{m}_{dq}^{c}$ are aligned with the converter domain, while $\Delta \boldsymbol{i}_{gdq}$, $\Delta \boldsymbol{u}_{gdq}$ and $\Delta \boldsymbol{m}_{dq}$ are aligned with the system domain. What's more, $\boldsymbol{I}_{gdq0}^{pll} = \begin{bmatrix} 0 & I_{gq0} T_{pll}(s) \\ 0 & -I_{gd0} T_{pll}(s) \end{bmatrix}$, $\boldsymbol{M}_{dq0}^{pll} = \begin{bmatrix} 0 & m_{q0} T_{pll}(s) \\ 0 & -m_{d0} T_{pll}(s) \end{bmatrix}$ and $\boldsymbol{U}_{gdq0}^{pll} = \begin{bmatrix} 1 & U_{gq0} T_{pll}(s) \\ 0 & 1 - U_{gd0} T_{pll}(s) \end{bmatrix}$.

*1.2.2 Inner current control loop*

The modulation ratio can be represented as:

$$\Delta \boldsymbol{m}_{dq}^{c} = -\frac{e^{-sT_s}}{V_{dc0}} \cdot \begin{bmatrix} \boldsymbol{H}_{cc} \left( \Delta \boldsymbol{i}_{gdq}^{c*} - \Delta \boldsymbol{i}_{gdq}^{c} \right) \\ -\boldsymbol{K}_d \Delta \boldsymbol{i}_{gdq}^{c} - \boldsymbol{H}_f \Delta \boldsymbol{u}_{gdq}^{c} \end{bmatrix} \qquad (6)$$

where $\boldsymbol{H}_{cc} = diag(H_{cc}(s) \ H_{cc}(s))$ and $\boldsymbol{H}_f = diag(H_f(s) \ H_f(s))$

By integrating the (5) into (6), the modulation ratio aligned with the system domain can be obtained as:

$$\Delta \boldsymbol{m}_{dq} = \underbrace{-\frac{e^{-sT_s}}{V_{dc0}} \boldsymbol{H}_{cc}}_{\boldsymbol{T}_{cc}^{1}} \Delta \boldsymbol{i}_{gdq}^{c*} + \underbrace{\frac{e^{-sT_s}}{V_{dc0}} \left( \boldsymbol{H}_{cc} + \boldsymbol{K}_d \right)}_{\boldsymbol{T}_{cc}^{2}} \Delta \boldsymbol{i}_{gdq} +$$
$$\underbrace{\left\{ \frac{e^{-sT_s}}{V_{dc0}} \left( \boldsymbol{H}_{cc} \boldsymbol{I}_{gdq0}^{pll} + \boldsymbol{K}_d \boldsymbol{I}_{gdq0}^{pll} + \boldsymbol{H}_f \boldsymbol{U}_{gdq0}^{pll} \right) - \boldsymbol{M}_{dq0}^{pll} \right\}}_{\boldsymbol{T}_{cc}^{3}} \Delta \boldsymbol{u}_{gdq} \qquad (7)$$

By integrating (7) into (2), the converter-side modulation voltage can be obtained as:

$$\Delta \boldsymbol{u}_{cdq} = \left( \boldsymbol{T}_{cc}^{1} \Delta \boldsymbol{i}_{gdq}^{c*} + \boldsymbol{T}_{cc}^{2} \Delta \boldsymbol{i}_{gdq} + \boldsymbol{T}_{cc}^{3} \Delta \boldsymbol{u}_{gdq} \right) V_{dc0} + \boldsymbol{m}_{dq0} \Delta u_{dc} \qquad (8)$$

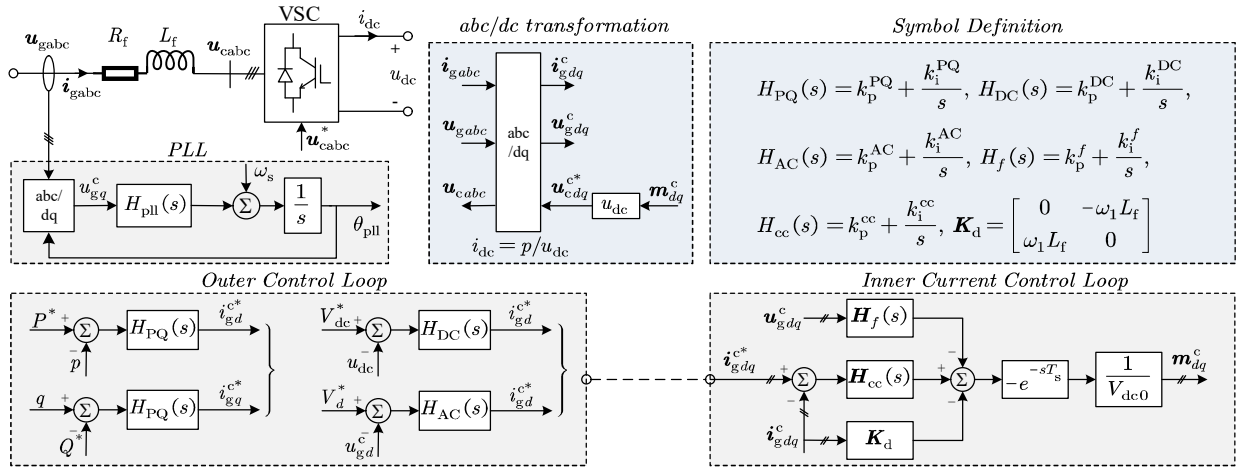

Fig. 1 Main circuit and control scheme of the VSC

## 1.2.3 Outer control loop

The active power and reactive power can be calculated as:

$$\Delta pq = \frac{3}{2}U_{gdq0}\Delta i_{gdq} + \frac{3}{2}I_{gdq0}\Delta u_{gdq} \quad (9)$$

where $\Delta pq = \begin{bmatrix} \Delta p \\ \Delta q \end{bmatrix}$, $U_{gdq0} = \begin{bmatrix} U_{gd0} & U_{gq0} \\ U_{gq0} & -U_{gd0} \end{bmatrix}$ and $I_{gdq0} = \begin{bmatrix} I_{gd0} & I_{gq0} \\ -I_{gq0} & I_{gd0} \end{bmatrix}$.

The alternating-power control and reactive-power control can be modelled as:

$$\Delta i_{gdq}^{c*} = \underbrace{\frac{3}{2}H_{PQ}U_{gdq0}}_{T_{PQ}^1}\Delta i_{gdq} + \underbrace{\frac{3}{2}H_{PQ}I_{gdq0}}_{T_{PQ}^2}\Delta u_{gdq} \quad (10)$$

where $H_{PQ} = diag(H_{PQ}(s)\ H_{PQ}(s))$.

The DC-voltage control and reactive power control can be modelled as:

$$\Delta i_{gdq}^{c*} = \underbrace{\frac{3}{2}H_{PQ}'U_{gdq0}}_{T_{DC}^1}\Delta i_{gdq}$$
$$+ \underbrace{\frac{3}{2}H_{PQ}'I_{gdq0}}_{T_{DC}^2}\Delta u_{gdq} + \underbrace{-H_{dc}'}_{T_{DC}^3}\Delta u_{dc} \quad (11)$$

where $H_{PQ}' = diag(0\ H_{PQ}(s))$, $H_{dc}' = [H_{dc}(s)\ 0]^T$.

The alternating-voltage control can be modelled as:

$$\Delta i_{gdq}^{c*} = \underbrace{H_{ac}U_{gdq0}^{pll}}_{T_{AC}^1}\Delta u_{gdq} \quad (12)$$

where $H_{ac} = diag(H_{ac}(s)\ H_{ac}(s))$.

## 1.3 Three-port admittance model

The three-port admittance models of three commonly-used operating modes are derived in the following part, including the P/Q mode, V/f mode and DC/Q mode.

### 1.3.1 P/Q mode

By integrating (10) into (8), the converter-side modulation voltage can be represented as:

$$\Delta u_{cdq} = (T_{cc}^1 T_{PQ}^1 + T_{cc}^2)V_{dc0}\Delta i_{gdq}$$
$$+ (T_{cc}^1 T_{PQ}^2 + T_{cc}^3)V_{dc0}\Delta u_{gdq} + m_{dq0}\Delta u_{dc} \quad (13)$$

By integrating (13) into (1), the ac-side admittance model can be obtained as:

$$\underbrace{\left(-Z_{dq}^f - (T_{cc}^1 T_{PQ}^1 + T_{cc}^2)V_{dc0}\right)}_{A}\Delta i_{gdq}$$
$$= \underbrace{\left((T_{cc}^1 T_{PQ}^2 + T_{cc}^3)V_{dc0} - I\right)}_{B}\Delta u_{gdq} + m_{dq0}\Delta u_{dc} \quad (14)$$

$$\Rightarrow \Delta i_{gdq} = \underbrace{A^{-1}B}_{Y_{dq}}\Delta u_{gdq} + \underbrace{A^{-1}m_{dq0}}_{a_{2\times 1}}\Delta u_{dc}$$

By integrating (1) into (3), the dc-side admittance model can be represented as:

$$\Delta i_{dc} = \underbrace{\frac{3}{2V_{dc0}}\left[(U_{cdq0}^T - I_{gdq0}^T Z_{dq}^f)Y_{dq} + I_{gdq0}^T\right]}_{b_{1\times 2}}\Delta u_{gdq}$$
$$+ \underbrace{\left[\frac{3}{2V_{dc0}}(U_{cdq0}^T - I_{gdq0}^T Z_{dq}^f)a_{2\times 1} - \frac{I_{dc0}}{V_{dc0}}\right]}_{Y_{dc}}\Delta u_{dc} \quad (15)$$

### 1.3.2 Vdc/Q mode

By integrating (11) into (8), the converter-side modulation voltage can be represented as:

$$\Delta u_{cdq} = (T_{cc}^1 T_{DC}^1 + T_{cc}^2)V_{dc0}\Delta i_{gdq} +$$
$$(T_{cc}^1 T_{DC}^2 + T_{cc}^3)V_{dc0}\Delta u_{gdq} + (T_{cc}^1 T_{DC}^3 V_{dc0} + m_{dq0})\Delta u_{dc} \quad (16)$$

By integrating (16) into (1), the ac-side admittance model can be obtained as:

$$\underbrace{\left(-Z_{dq}^f - (T_{cc}^1 T_{DC}^1 + T_{cc}^2)V_{dc0}\right)}_{A}\Delta i_{gdq} =$$
$$\underbrace{\left[(T_{cc}^1 T_{DC}^2 + T_{cc}^3)V_{dc0} - I\right]}_{B}\Delta u_{gdq} + (T_{cc}^1 T_{DC}^3 V_{dc0} + m_{dq0})\Delta u_{dc}$$

$$\Rightarrow \Delta i_{gdq} = \underbrace{A^{-1}B}_{Y_{dq}}\Delta u_{gdq} + \underbrace{A^{-1}(T_{cc}^1 T_{DC}^3 V_{dc0} + m_{dq0})}_{a_{2\times 1}}\Delta u_{dc}$$
$$(17)$$

The dc-side admittance model is the same as the (15) except for substituting the $Y_{dq}$ and $a_{2\times 1}$ calculated in (17).

### 1.3.3 V/f mode

Under *V/f* mode, the system frequency is given by the converter, meaning that PLL dynamics can be ignored. The modulation ratio model in (7) can be simplified as:

$$\Delta m_{dq} = \underbrace{-\frac{e^{-sT_s}}{V_{dc0}}H_{cc}}_{T_{cc}^{1'}}\Delta i_{gdq}^* + \underbrace{\frac{e^{-sT_s}}{V_{dc0}}(H_{cc} + K_d)}_{T_{cc}^{2'}}\Delta i_{gdq} \quad (18)$$

By integrating (18) into (8), the converter-side modulation voltage can be represented as:

$$\Delta u_{cdq} = (T_{cc}^{1'}\Delta i_{gdq}^* + T_{cc}^{2'}\Delta i_{gdq})V_{dc0} + m_{dq0}\Delta u_{dc} \quad (19)$$

The alternating-voltage control in (12) can be simplified as:

$$\Delta \boldsymbol{i}_{\mathrm{g}dq}^{*} = \boldsymbol{H}_{\mathrm{ac}}\Delta \boldsymbol{u}_{\mathrm{g}dq} \quad (20)$$

By integrating (20) into (19), the converter-side modulation voltage can be represented as:

$$\Delta \boldsymbol{u}_{\mathrm{c}dq}=\left(\boldsymbol{T}_{\mathrm{cc}}^{1'}\boldsymbol{H}_{\mathrm{ac}}\Delta \boldsymbol{u}_{\mathrm{g}dq}+\boldsymbol{T}_{\mathrm{cc}}^{2'}\Delta \boldsymbol{i}_{\mathrm{g}dq}\right)V_{\mathrm{dc}0}+\boldsymbol{m}_{dq0}\Delta u_{\mathrm{dc}} \quad (21)$$

By integrating (21) into (1), the ac-side admittance model can be represented as:

$$\underbrace{\left(-\boldsymbol{Z}_{dq}^{\mathrm{f}} - \boldsymbol{T}_{\mathrm{cc}}^{2'}V_{\mathrm{dc}0}\right)}_{\boldsymbol{A}}\Delta \boldsymbol{i}_{\mathrm{g}dq} =$$
$$\underbrace{\left[\boldsymbol{T}_{\mathrm{cc}}^{1'}\boldsymbol{H}_{\mathrm{ac}}V_{\mathrm{dc}0} - \boldsymbol{I}\right]}_{\boldsymbol{B}}\Delta \boldsymbol{u}_{\mathrm{g}dq} + \boldsymbol{m}_{dq0}\Delta u_{\mathrm{dc}} \quad (22)$$
$$\Rightarrow \Delta \boldsymbol{i}_{\mathrm{g}dq} = \underbrace{\boldsymbol{A}^{-1}\boldsymbol{B}}_{\boldsymbol{Y}_{dq}}\Delta \boldsymbol{u}_{\mathrm{g}dq} + \underbrace{\boldsymbol{A}^{-1}\boldsymbol{m}_{dq0}}_{\boldsymbol{a}_{2\times1}}\Delta u_{\mathrm{dc}}$$

The dc-side admittance model is the same as the (15) except for substituting the $\boldsymbol{Y}_{dq}$ and $\boldsymbol{a}_{2\times 1}$ calculated in (22). Based on the (14), (15), (17) and (22), the three-port admittance of the VSC can be represented in complex vector as:

$$\begin{bmatrix}\Delta \boldsymbol{i}_{\mathrm{g}dq}\\ \Delta i_{\mathrm{dc}}\end{bmatrix} = \begin{bmatrix}\boldsymbol{Y}_{dq} & \boldsymbol{a}_{2\times 1}\\ \boldsymbol{b}_{1\times 2} & Y_{\mathrm{dc}}\end{bmatrix}\cdot\begin{bmatrix}\Delta \boldsymbol{u}_{\mathrm{g}dq}\\ \Delta u_{\mathrm{dc}}\end{bmatrix} \quad (23)$$

## 3 Model Validation

### 3.1 Frequency scanning method

The method for obtaining one-dimensional and two-dimensional admittances from simulations has been well discussed [3]. Here, the frequency scanning method for obtaining the above three-dimensional model is introduced. As shown Fig.2, the first step is to select a vector of frequencies $f_{dq,\mathrm{tab}}$, i.e., the frequencies at which the admittances shall be calculated. Taking the series voltage injection for example, three perturbation signals will be injected from ac- and dc-side.

$$\mathrm{AC}: v_{inj1} = V_{inj}\begin{bmatrix}\sin\left(\left[\omega_{inj}+\omega_1\right]t\right)\\ \sin\left(\left[\omega_{inj}+\omega_1\right]t\right)-2\pi/3\\ \sin\left(\left[\omega_{inj}+\omega_1\right]t\right)+2\pi/3\end{bmatrix}$$

$$\mathrm{AC}: v_{inj2} = V_{inj}\begin{bmatrix}\sin\left(\left[\omega_{inj}-\omega_1\right]t\right)\\ \sin\left(\left[\omega_{inj}-\omega_1\right]t\right)+2\pi/3\\ \sin\left(\left[\omega_{inj}-\omega_1\right]t\right)-2\pi/3\end{bmatrix} \quad (24)$$

$$\mathrm{DC}: v_{inj3} = 2V_{inj}\left[\sin\left(\left[\omega_{inj}\right]t\right)\right]$$

After converting time-domain signals to the frequency domain as described in Fig. 2, the following equations can be used to find the three-port admittances:

$$\begin{bmatrix}Y_{dd} & Y_{dq} & a_1\\ Y_{qd} & Y_{qq} & a_2\\ b_1 & b_2 & Y_{dc}\end{bmatrix} = \begin{bmatrix}v_{d1} & v_{d2} & v_{d3}\\ v_{q1} & v_{q2} & v_{q3}\\ v_{dc1} & v_{dc2} & v_{dc3}\end{bmatrix}\begin{bmatrix}i_{d1} & i_{d2} & i_{d3}\\ i_{q1} & i_{q2} & i_{q3}\\ i_{dc1} & i_{dc2} & i_{dc3}\end{bmatrix}^{-1} \quad (25)$$

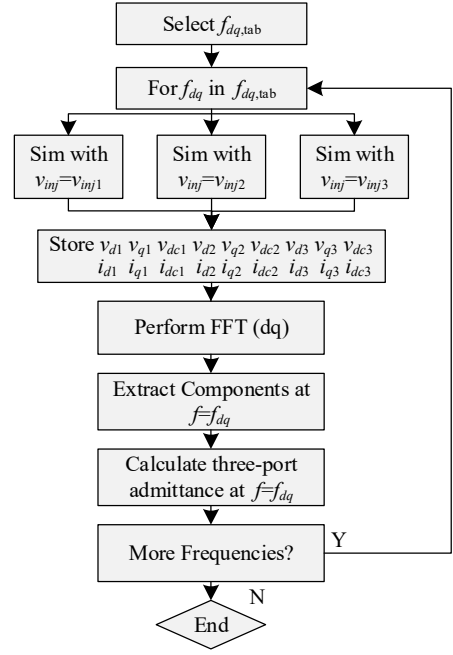

Fig. 2 The frequency scanning flowchart

### 3.2 Model validation

The parameters of VSC under three commonly-used control schemes are given in Table 1. The corresponding simulation model is built in the Matlab/Simulink.

Table 1 VSC parameters

| Mode | Parameters | Data |
| --- | --- | --- |
| VSC (PQ) | Bandwidth (PQ ; PLL;CC) | 2.5;40;200 (Hz) |
| | ac-side filter | 66.7μΩ; 2μH |
| | dc-side capacitor; control delay | 20mF; 100ms |
| | power rating (active; reactive) | 300MW; 0Mvar |
| VSC (V/f) | control bandwidth (AC; CC) | 25; 200 (Hz) |
| | ac-side filter | 1mΩ; 66.5 mH |
| | dc-side capacitor; control delay | 22.5μF; 100ms |
| VSC (DC/Q) | Bandwidth (PQ; DC; PLL; CC) | 3;6;40;200 (Hz) |
| | ac-side filter | 1mΩ; 66.5 mH |
| | dc-side capacitor; control delay | 22.5μF; 100ms |
| | power rating (reactive) | 0 Mvar |

The comparisons between the frequency scanning results and the theoretical three-port modelling are presented in Fig. 1. From which, it can be seen that the accuracy of theoretical models is high, which validates the correctness of the established three-port admittance model.

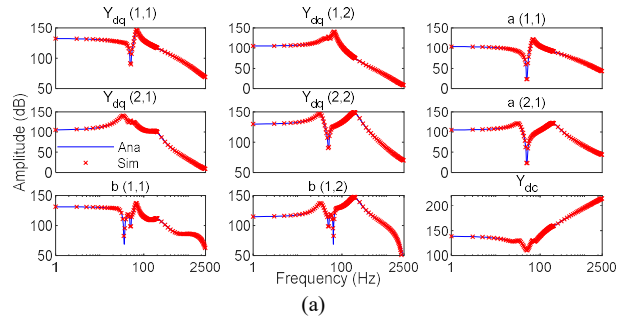

(a)

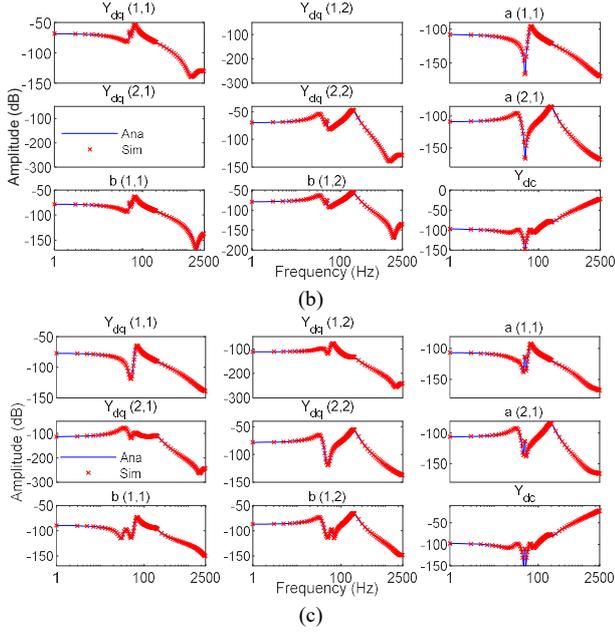

Fig. 3 Model validation, (a) PQ; (b) V/f; (c) DC/Q

## 4 Stability analysis

In this section, the stability analysis of a typical AC/DC converter system is given using the above established three-port admittance model. The structure of the studied system is given in Fig. 4, where the converter's ac-side is connected to an unideal ac-grid, and its dc-side is connected with a dc transmission network.

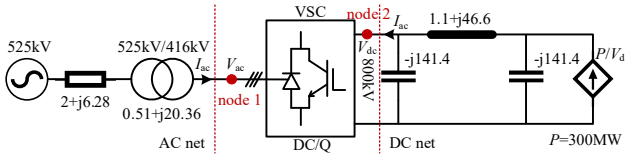

Fig. 4 The topology of the studied AC/DC system

*4.1 Theoretical Analysis*

Two commonly-used stability analysis methods are both applied in this part. The first one utilizes the system open-loop model and apply the generalized Nyquist criterion (GNC) for the stability analysis. The second one utilizes the system closed-loop model and calculated the pole-zero distribution for the stability analysis.

*4.1.1 Open-loop model-based stability analysis*

As shown in Fig. 4, the open-loop impedance models of the AC network and DC network can be represented as:

$$\begin{bmatrix} V_{acd} \\ V_{acq} \\ V_{dc} \end{bmatrix} = \underbrace{\begin{bmatrix} Z_{dd}^{ac} & Z_{dq}^{ac} & 0 \\ Z_{qd}^{ac} & Z_{qq}^{ac} & 0 \\ 0 & 0 & Z^{dc} \end{bmatrix}}_{\boldsymbol{Z}_{\text{net}}} \begin{bmatrix} I_{acd} \\ I_{acq} \\ I_{dc} \end{bmatrix} \quad (26)$$

The open-loop admittance model of the VSC is like:

$$\begin{bmatrix} I_{acd} \\ I_{acq} \\ I_{dc} \end{bmatrix} = \underbrace{\begin{bmatrix} Y_{dd} & Y_{dq} & a_1 \\ Y_{qd} & Y_{qq} & a_2 \\ b_1 & b_2 & Y_{dc} \end{bmatrix}}_{\boldsymbol{Y}_{\text{con}}} \begin{bmatrix} V_{acd} \\ V_{acq} \\ V_{dc} \end{bmatrix} \quad (27)$$

According to [5], the impedance ratio of the studied system can be constructed as $\boldsymbol{Z}_{\text{net}}\boldsymbol{Y}_{\text{con}}$, whose stability can be judged via the GNC. The marginal unstable condition is set as: increasing the internal impedance of the ac-grid from 2+$j$6.28 (20mH) from 2+$j$9.42 (30mH). The root locus of the $\det(\boldsymbol{Z}_{\text{net}}\boldsymbol{Y}_{\text{vsc}})$ is shown in Fig. 5. From which, the theoretical results (surround the point (-1, $j$0)) indicate that the system is unstable with internal impedance equals to 2+$j$9.42.

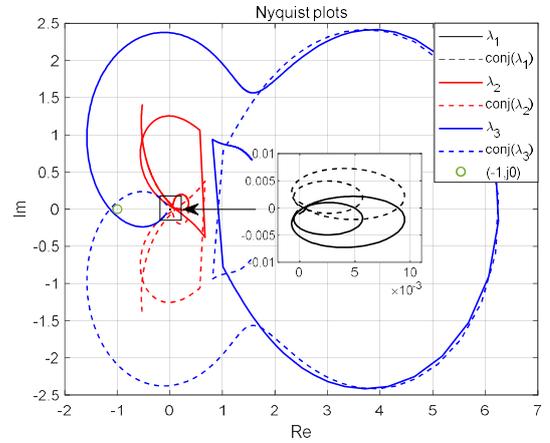

Fig. 5 Open-loop method-based stability analysis

*4.1.2 Closed-loop model-based stability analysis*

The closed loop model of the studied system can be obtained by formulating the system nodal admittance matrix. As can be seen from the Fig. 4, the studied system constitutes of two nodes, which can be modelled as:

$$\boldsymbol{Y}_{\text{node}} = \begin{bmatrix} Y_{dd} + Y_{dd}^{ac} & Y_{dq} + Y_{dq}^{ac} & a_1 \\ Y_{qd} + Y_{qd}^{ac} & Y_{qq} + Y_{qq}^{ac} & a_1 \\ b_1 & b_2 & Y^{dc} + Y_{dc} \end{bmatrix} \quad (28)$$

According to [11], the stability can be judged by calculating the equation of $\det(\boldsymbol{Y}_{\text{node}}) = 0$. The system is stable if and only if there is no RHP-zeros (right-half-plane). The marginal unstable condition is set as the same as the 4.1.1. Fig. 6 presents the system pole-zero distribution. From which, it can be seen that there appear one pair of unstable RHP-zeros with internal impedance equals to 2+$j$9.42.

*4.2 Simulation Validations*

From the above analysis, it can be seen that the open-loop-based and the closed-loop-based stability analysis will give the same stability conclusion that the system should be unstable when the ac-grid internal inductance equals to 30mH. The corresponding time-domain simulations are presented in Fig. 7, which are consistent with the theoretical results. The effectiveness of the three-port admittance model-based AC/DC stability analysis has been verified.

## 5 Conclusion

In this paper, the three-port admittance model of the VSC is developed and its accuracy is validated by the proposed measurement method from simulations. Based on such model, two stability analysis methods (i.e., open-loop and closed loop) are both applied in the converter's AC/DC stability evaluation. It is found that the above two methods can obtain the identical stability conclusion. In which, the open-loop method relies on the inspection of the Nyquist root locus, while the closed-loop method is based on the calculation of RHP poles-zeros. In the authors' future work, more complicated AC/DC interconnected system will be considered to broaden the application scope of the three-port admittance model-based stability analysis method.

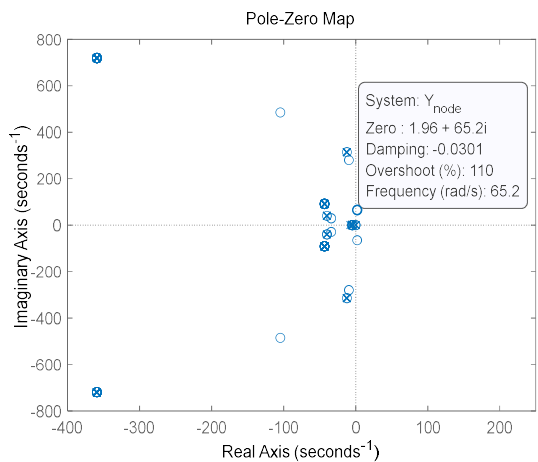

Fig. 6 Closed-loop method-based stability analysis

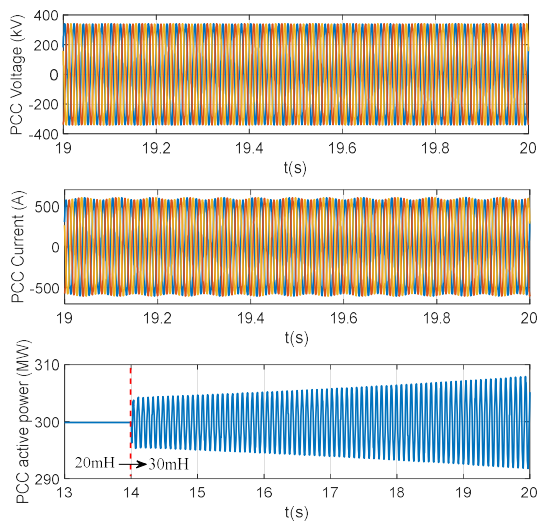

Fig. 7 Time domain simulation

## 6 Acknowledgements

This work was supported by the National Natural Science Foundation of China under Grant 51837007. The authors would like to thank team members at SJTU, and co-workers from NTNU for their valuable advices.